\documentstyle[bezier,12pt]{article}
\newcommand{\bra}[1]{\langle #1 |}
\newcommand{\ket}[1]{| #1 \rangle}

\newcommand{\eq}[1]{eq.(\ref{#1})}
\newcommand{\dpar}[2]{\frac{\partial #1}{\partial #2}}

\newcommand{\bm}[1]{\mbox{\boldmath $#1$}}
\def\ltap{\raisebox{-.55ex}{\rlap{$\sim$}} \raisebox{.4ex}{$<$}}
\def\gtap{\raisebox{-.55ex}{\rlap{$\sim$}} \raisebox{.4ex}{$>$}}
\def\gsim{\mathrel{\gtap}}
\def\lsim{\mathrel{\ltap}}

\def\e{\mbox{e}}

\def\Re{\mathop{\mbox{Re}}}
\def\Im{\mathop{\mbox{Im}}}
\def\half{{1 \over 2}}

\begin{document}

\title{Numerical Study of Induced False Vacuum Decay at High Energies}

\author{
  A.N.Kuznetsov and P.G.Tinyakov \\
  {\small {\em Institute for Nuclear Research of the Russian
   Academy of Sciences,}}\\
  {\small {\em 60th October Anniversary prospect,
   7a, Moscow 117312, Russia.}}
  }
\date{October 1995}
\maketitle
\begin{abstract}
We calculate numerically the probability $\exp[ {1\over\lambda}
F(E/E_{sph},N/N_{sph})]$ of the false vacuum decay in the massive
four-dimensional $-\lambda\phi^4$ model from multiparticle initial
states with fixed number of particles $N$ and energy $E$ greater than
the height of the barrier $E_{sph}$. We find that at $E\lsim 3E_{sph}$
and $N\lsim 0.4N_{sph}$ the decay is classically forbidden and thus is
exponentially suppressed. We argue that the classically forbidden
region extends at small $N$
at least up to $E\sim 10 E_{sph}$ and, most likely, to
all energies. Our data suggest that the false vacuum decay induced by
two-particle collisions is exponentially suppressed at all
energies.
\end{abstract}

\newpage

\section{Introduction}

In the last few years a substantial progress has been made in
understanding of non-vacuum tunneling in field theory. The interest to
this subject is inspired by attempts to calculate, in the Standard
Model, the instanton-mediated baryon number violation in particle
collisions at high energy~\cite{RE}. Although a variety of methods has
been developed for calculating the tunneling probabilities at $E\ll
E_{sph}$, where $E_{sph}$ is the height of the barrier (the
sphaleron energy in the Electroweak Theory), the behaviour of
probabilities in the most interesting region $E \gsim E_{sph}$ is
still unknown (for a review see ref.\cite{reviews}). In particular,
the question of whether the exponential suppression of baryon number
violating processes disappears at some sufficiently high energy of
colliding particles, still remains unanswered. Since analytical
approaches to tunneling induced by particle collisions
seem to be exhausted, in the
present paper we address this problem by means of numerical methods.

A suitable starting point for numerical study of non-vacuum tunneling
is provided by the semiclassical approach developed in
refs.\cite{RT,T,RSTboundary}. In this approach the key role is played by the
probability of tunneling from a mixed state with fixed energy $E$ and
number of particles $N$. This probability is defined as follows,
\[
  \sigma(E,N) = \sum_{i,f} |\bra{f} \hat{S} \hat{P}_E \hat{P}_N \ket{i}|^2,
\]
where $\hat{S}$ is the S-matrix, $\hat{P}_{E,N}$ are projectors onto
subspaces of fixed energy $E$ and fixed number of particles $N$,
respectively, while the states $\ket{i}$ and $\ket{f}$ are
perturbative excitations above two vacua lying on different sides of
the barrier. It was argued in refs.\cite{RT,T} that at any fixed $N$,
the probability $\sigma(E,N)$ sets an upper bound for the two-particle
cross section, while in the limit of small $N$ this probability
reproduces the two-particle cross section with exponential accuracy.

The advantage of considering multiparticle probability $\sigma(E,N)$
instead of two-particle one is that in the weak coupling regime
$g^2\to 0$ and $E,N \sim 1/g^2$, the former can be calculated
semiclassically and has the form
\begin{equation}
  \sigma(E,N) \sim \exp \Bigl\{ {1 \over g^2} F(\epsilon,\nu)
  \Bigr\} ,
\label{general form}
\end{equation}
where $\epsilon = E/E_{sph}$, $\nu = N/N_{sph}$ and $N_{sph}\sim
1/g^2$ is the number of particles produced in the sphaleron decay. The
function $F(\epsilon,\nu)$ is negative at low energies, which
corresponds to exponential suppression of probability in this
domain. At $\epsilon =1$ and $\nu=1$ (at the sphaleron), the function
$F(\epsilon,\nu)$ vanishes, i.e. the exponential suppression disappears.
The absence of exponential suppression of the two-particle cross
section would show up as zero of the function $F(\epsilon,\nu)$ at
some fixed $\epsilon > 1$ and $\nu\to 0$.

The function $F(\epsilon,\nu)$ is determined by a solution to the
specific classical boundary value problem for the complexified field
equations~\cite{RSTboundary}. Namely, one has to solve the usual field
equation
\begin{equation}
  {\delta S\over\delta \phi} =0,
\label{field eq}
\end{equation}
where time and the field $\phi$ are treated as complex variables.  All
the information about particular problem is encoded into the boundary
conditions, which are formulated on the contour ABCD in the complex
time plane (see Fig.1).  The asymptotic regions A and D correspond to
the initial and final states, respectively.

The boundary conditions are as follows. The final boundary condition
states that the field is real on the line CD, where it represents the
classical evolution of the system after barrier penetration.
This requirement can be imposed at the point C
where it implies
\[
  \Im \phi = 0,
\]
\begin{equation}
  \Im \dpar{\phi}{t} =  0.
\label{bc 1}
\end{equation}
The part CD of the contour is not essential for formulation of these
boundary conditions. In fact, this part does not play any role in the
calculation of the function $F(\epsilon,\nu)$ and can be dropped,
unless the details of the final state are of interest.  Due to this
fact the final boundary conditions (\ref{bc 1}) apply also to
$-\lambda\phi^4$ model where the stable vacuum does not exist and
field develops a singularity in the final state, i.e. somewhere on the
line CD (point P in Fig.1).

At the other end of the contour, in the asymptotic region $A$,
the field is required to be linear. The initial boundary conditions
fix the ratio of amplitudes in the negative- and positive-frequency
parts to be a constant independent of momentum. If one writes
general asymptotics of the field in the form
\begin{equation}
  \phi(x) = \int { d{\bm k} \over \sqrt{(2\pi)^3 2\omega_k} }
  \Bigl\{ \e^{-\theta} f_{\bf k} \e^{-i\omega_k\eta+i{\bf kx}}
  + g_{\bf k} \e^{i\omega_k\eta-i{\bf kx}}  \Bigr\} ,
\label{asymptotic form}
\end{equation}
where $\eta = \Re t$ and $\theta$ is a real positive parameter, then the
boundary condition is
\begin{equation}
  g_{\bf k}=(f_{\bf k})^*.
\label{bc 2}
\end{equation}
This equation fixes one particular linear combination of the field and
its time derivative. In total, there are two real conditions at
each end of the contour for the second order complex differential
equation, so in general the boundary value problem is completely
specified.

The input variables $E$ and $N$ enter the above boundary conditions
through two parameters: $\theta$, which enters the boundary conditions
explicitly, and $T$, which is the amount of Euclidean evolution (see
Fig.1) and specifies the place where initial boundary conditions are
imposed. The relation between $T$, $\theta$ and $E$, $N$ is given by
the equations
\[
  E = - \dpar{}{T}\Re[iS(T,\theta)],
\]
\begin{equation}
  N = - 2\dpar{}{\theta}\Re[iS(T,\theta)],
\label{EN eqs}
\end{equation}
where $S(T,\theta)$ is the action of the solution for given $T$ and
$\theta$, evaluated along the contour.  Note that according to
boundary conditions (\ref{bc 1}) the field is real on the line CD, so
that this part of the contour does not contribute into
$\Re[iS(T,\theta)]$.  Alternatively, the energy and the number of
particles can be read off from the initial asymptotics of the
field,
\[
  N = \int d{\bm k} f_{\bf k} f^*_{\bf k} \, ,
\]
\begin{equation}
  E = \int d{\bm k} \omega_{\bf k} f_{\bf k} f^*_{\bf k} \, .
\label{ENspec}
\end{equation}
One can check that eqs.(\ref{EN eqs}) and (\ref{ENspec})
coincide, provided that eqs.(\ref{field eq}) and (\ref{bc 2}) are
satisfied.

Given the solution to the boundary value problem one can
calculate the function $F(\epsilon,\nu)$ according to the formula
\begin{equation}
  {1\over g^2}F(\epsilon,\nu)
  = 2ET + N\theta + 2\Re [iS(T,\theta)]\, ,
\label{F=}
\end{equation}
where $T$ and $\theta$ depend on $E$ and $N$ through
eqs.(\ref{EN eqs}).

Several remarks are in order. First, as follows from
eqs.(\ref{asymptotic form}) and (\ref{bc 2}), at $\theta\neq 0$ the
solution is necessarily complex along the line AB. Thus, contributions
to the function $F(\epsilon,\nu)$ come from both Minkowskian (AB) and
Euclidean (BC) parts of the contour. Second, as the field equations
are analytic in time, the solution is also analytic everywhere except
possible singularities. Thus, the contour ABC can be deformed,
provided that the asymptotic region A is untouched and singularities
are not crossed. In fact, one {\it must} expect singularities in
between the contour ABC and the negative part of the real time
axis. Otherwise the reality of the field at real $t$ and the initial
boundary conditions are incompatible. In the model we consider below
these singularities lie on the real time axis, as shown in Fig.1. In
numerical calculations it is convenient to deform the contour in such
a way that it passes far from singularities (dotted line in
Fig.1).

In one particular case, namely at $\theta=0$, the above boundary value
problem simplifies considerably. In this case the initial boundary conditions,
\eq{bc 2}, reduce to the reality condition imposed at $\Im t = T$. The
solution to the resulting boundary value problem is given by periodic
instanton of ref.\cite{KRTperiod}. Periodic instanton is a real
periodic solution to the Euclidean field equations with period $2T$
and two turning points at $t=0$ and $t=iT$. Being analytically
continued to the Minkowskian domain through the turning points,
periodic instanton stays real at lines $\Im t =0$ and $\Im t = T$ and
thus satisfies the boundary conditions with $\theta=0$.

The periodic instanton can be found analytically in two extreme cases:
at $\epsilon \ll 1$ it can be approximated by the periodic chain of
instantons and antiinstantons, while at $1-\epsilon \ll 1$ it is
approximately given by the oscillations in the sphaleron negative
mode. At intermediate energies the periodic instanton can be obtained
numerically~\cite{Matveevml,HMT}. In the $\epsilon$--$\nu$ plane, periodic
instantons form a line connecting the points $\epsilon=\nu=0$ (zero
energy instanton) and $\epsilon=\nu=1$ (sphaleron). The function
$F(\epsilon,\nu)$ monotonically grows from $F(0,0)=-S_{inst}$ to
$F(1,1)=0$ along this line.

The rest of this paper is devoted to numerical solution of the
boundary value problem specified by eqs.(\ref{field eq}), (\ref{bc 1})
and (\ref{bc 2}) in the general case $\theta \neq 0$, and
investigation of the behaviour of the function $F(\epsilon,\nu)$ at
$\epsilon \gsim 1$. For simplicity reasons we concentrate on
particle-induced false vacuum decay in the theory with one scalar
field.

\section{The Model}

The choice of particular model turns out to be strongly constrained for
technical reasons.  After several attempts we ended up with
four-dimensional $-\lambda\phi^4$ theory with the mass term. The
action of this model reads
\begin{equation}
  S = \int d^4 x \Bigl( \half \partial_{\mu} \phi\partial^{\mu} \phi
  - \half m^2\phi^2   + {1\over 4}\lambda\phi^4 \Bigr),
\label{action}
\end{equation}
where $\lambda$ is a positive constant.  At non-zero $m$ the region
$\phi \approx 0$ and the instability region $\phi >
m/\sqrt{\lambda}$ are separated by a finite energy barrier,
and one can ask whether the presence of colliding particles enhances
the probability of the barrier penetration (i.e., the decay of the
metastable state $\phi=0$).

It is a lucky coincidence that this model is of special interest for
another reason, as it has many features reminiscent of bosonic sector
of the Electroweak Theory. At $m=0$ (zero Higgs vacuum expectation
value) both models are conformally invariant and possess instanton
solutions. In the case of $-\lambda\phi^4$ model the
instanton  \cite{FL} has the action
\begin{equation}
  S_{inst} = {8 \pi^2 \over 3\lambda}
\label{Sinst}
\end{equation}
and describes the decay of the metastable state $\phi=0$.  At $m\neq
0$ (non-zero Higgs vacuum expectation value) the conformal symmetry is
softly broken and instanton solutions disappear, while the low energy
transitions are described by constrained instantons~\cite{Affleck}.
The action (\ref{Sinst}) gets small energy-dependent
correction and still determines the transition probability.  Like the
Electroweak Theory, the $-\lambda\phi^4$ model possesses the sphaleron
solution. The energy of the sphaleron can be found numerically,
\begin{equation}
  E_{sph} = 18.9 {m\over\lambda}.
\label{Esph}
\end{equation}

Before discussing the discrete formulation of the boundary value
problem specified above, it is convenient to rewrite the action
(\ref{action}) in the dimensionless variables. Since the problem is
O(3)-symmetric, we restrict ourselves to $s$-wave scattering. The
change of variables
\[
\phi = {1\over |{\bm x}| \sqrt{\lambda} }\psi,
\]
\[
x = m^{-1}y,
\]
brings the action for spherically symmetric configurations to the form
\begin{equation}
S = {4 \pi \over\lambda} \int dt\int\limits_0^{\infty}dr \Bigl[
\half (\partial_t \psi)^2
- \half (\partial_r \psi)^2
- \half \psi^2   + {1\over 4r^2}\psi^4 \Bigr],
\label{action2}
\end{equation}
where $r = |{\bm y}|$ is the dimensionless radial distance. Throughout
the rest of this paper all dimensionfull quantities are measured in the
units of mass.

The starting point for our calculations is the discretized
version of the action (\ref{action2}).
The system is put on a grid of the size $L$
with $n_x+1$ sites $r_j = jL/n_x$, $j=0\ldots n_x$. Similarly, the
time contour is represented by the set of complex points $t_i$ with
$i=0\ldots n_t$. The field $\psi(t,r)$ transforms into the set
of complex variables $\psi_{ij}$. To define integrations, we introduce
two sets of intervals for each coordinate,
\[
dr_j = r_{j+1} - r_j, \hskip 0.5cm j=0\ldots n_x-1,
\]
\[
\tilde{dr_j} = (dr_{j-1} + dr_j)/2, \hskip 0.5cm j=1\ldots n_x-1,
\]
\[
\tilde{dr}_{0,n_x} = dr_{0,n_x}/2 ,
\]
and similarly for $dt_i$ and $\tilde{dt}_i$ (tilted and non-tilted
intervals are used to integrate fields and derivatives, respectively).
With these definitions, the discretized action reads
\begin{equation}
S = {4 \pi \over\lambda} \sum\limits_{ij} \Bigl[
\half (\psi_{i+1,j}-\psi_{ij})^2 {\tilde{dr_j}\over dt_i}
- \half (\psi_{i,j+1}-\psi_{ij})^2 {\tilde{dt_i}\over dr_j}
- V_{ij}\tilde{dt_i}\tilde{dr_j}\Bigr],
\label{discrete action}
\end{equation}
where
\[
V_{ij} = \half \psi_{ij}^2 -  {1\over 4r_j^2}\psi_{ij}^4 \;\;\; \mbox{at}
\;\;\; j\neq 0,
\]
\[
V_{i0}=0\, .
\]
Equations of motion can be easily derived from this action by
variation with respect to $\psi_{ij}$.

We also need the discrete version of the boundary conditions.
Let us start with the final boundary conditions. The first
equation (\ref{bc 1}) transforms into
\[
\Im \psi_{n_t,j} = 0\; .
\]
The discrete analog of the second equation can be derived from the
requirement that the solution, being continued to the real time axis
according to the {\it discretized} equations of motion, stays real.
This requirement leads to the equations
\[
 \Im \dpar{S}{\psi_{n_t,j}} = 0.
\]

The discrete reformulation of the initial boundary conditions requires
somewhat more work. It can be obtained by means of the following
trick. Let us make one step back and treat time as continuous
variable. Then we would obtain the $n_x$-dimensional quantum
mechanical system with the action trivially derived from \eq{discrete
action}.  Since it is assumed that the field reaches linear regime in
the initial asymptotic region, we can restrict ourselves to the
quadratic part of the resulting action and diagonalize it
numerically. In this way we obtain the discrete analog of the momentum
representation.  Given the corresponding eigenvalues and eigenvectors, it
is straightforward to perform the decomposition of the field into
positive- and negative-frequency parts and impose boundary
conditions~(\ref{bc 2}). Eqs.(\ref{bc 2}) are then translated into conditions
imposed on some linear combinations of $\psi_{0j}$ and $\psi_{1j}$.
These are the desired initial boundary conditions in the discrete
formulation. To save space we do not present the corresponding
cumbersome expressions.

It is instructive to note that the above $n_x$-dimensional quantum
mechanical system itself can be viewed as a model for studying
tunneling transitions from the excited states in multidimensional
systems. The action (\ref{discrete action}) can be treated as
discretized (in time) version of the action for this quantum
mechanical system. In practice this means that particular
number of lattice sites
in space direction is not crucial for our conclusions, as long as
linear regime can be reached at the initial part of the contour.

\section{Numerical method}

To solve the boundary value problem for given values of $T$ and
$\theta$ we use a multidimensional version of Newton's
method. This is a relaxation procedure which takes as input an
approximate solution to the non-linear field equations and improves it
at each iteration by solving the linearized equations in the
background of the current approximation. Iterations are repeated until
non-linear equations are satisfied to the desired accuracy.  The
advantage of this algorithm is that its convergency requires neither
positive-definiteness nor even reality of the action.  Moreover, the
convergency is quadratic, provided that initial approximation is
choosen sufficiently close to the solution. In practice, the accuracy
of $10^{-10}$ is reached in 3--6 iterations.

A drawback of the Newton's method is that its basin of convergency can
be very narrow. Thus, it is important to have a good initial
approximation at least for some values of $T$ and $\theta$.  Then one
can move gradually in the $T$ -- $\theta$ plane using output of each
run as input for the next one. In our case the starting configuration
is provided by the periodic instanton which, at $E\approx E_{sph}$,
can be approximated by sphaleron plus harmonic oscillation in the
sphaleron negative mode (both the sphaleron and its negative
mode have to be found numerically). The period of the periodic
instanton at $E\to E_{sph}$ is determined by the sphaleron negative
eigenvalue and equals
\[
  T_{crit} = 0.78\; .
\]
The amplitude of the oscillation goes to zero when $T\to T_{crit}$.
So, we take $T\approx T_{crit}$, $\theta=0$, and adjust the amplitude
of the oscillation in order to get inside the basin of convergency of
the periodic instanton with the period $T$. In this way we obtain the
very first solution to our boundary value problem.  Changing $T$ by
small steps we then reproduce the whole family of periodic instantons
with different periods. Calculations show that in the $-\lambda\phi^4$
model the period of the periodic instanton varies from $T_{crit}$ to
zero when the energy decreases from $E_{sph}$ to zero. This unusual
behaviour of period with energy is much in common with that in the
O(3) sigma model \cite{HMT}, where it is also related to
softly broken conformal invariance. On the basis of this similarity
one should expect the analogous behaviour in the Electroweak
Theory.

The Newton's method reduces the non-linear boundary value problem to
sequential solution of a few {\em linear} boundary value problems.  At
each iteration, to find the correction $\delta\psi$ to the current
approximation amounts to one matrix inversion,
\[
  \delta\psi = D^{-1}R \; ,
\]
where $D$ is a matrix of second derivatives of the action and $R$ is a
vector of first derivatives, both evaluated at the background.  The
matrix $D$ has the dimension $(2n_t n_x)\times(2n_t n_x)$, the factor
2 being due to the complexity of the field. In general, the computer
time necessary for inversion of such a matrix scales like $(n_t
n_x)^3$.  One can use, however, the fact that this matrix originates
from the local second order differential equation and thus is
sparse. The sparseness enables one to invert this matrix in $\propto
n_t n_x^3$ operations. Note that for systems containing $n_f$ fields
instead of one, this number would be $n_t (n_fn_x)^3$. In practice, to
find one solution to the boundary value problem at the grid of
dimension $n_t\times n_x = 200\times 40$ takes about 5 minutes at
SPARCstation 20, most of this time being spent for matrix
inversion. The necessary amount of memory scales like $n_tn_x^2$ and
for the above grid is of order 10 Mb.

Let us now discuss constraints on other grid parameters. The most
restrictive requirement is that the field must reach linear regime in
the asymptotic region A. For that the number of independent modes
(which equals $n_x$) must be large enough, and the length of Minkowskian
part of the contour, $T_M$, must be sufficient to let the energy
spread over these modes. Since in the Minkowskian region the most of
the field propagates along the light cone, the space size $L$ should
be larger than $T_M$ to avoid boundary effects.  On the other hand,
the lattice spacing in $r$ direction, $\Delta r$, should be small enough not
only to be close to the continuum limit, but also because
otherwise the free spectrum would be cut at too low frequency,
$\omega_{max} \sim \pi/\Delta r$. This would impose constraints on
the available region in the $E$--$N$ space, $E/N \ll \omega_{max}$, and
would not allow for the initial states consisting of small number of
high energy particles.  Thus, all these arguments push towards large
$n_x$, which is however bound at relatively low value $n_x\sim 50-100$
by the abilities of available computers.

The number of time slices $n_t$ is not so constrained since the
required computer resources scale linearly with $n_t$. In our
calculations we take $n_t$ few times larger than~$n_x$ in order to
saturate the continuum limit in time. This number is different for
different $T$ and $\theta$.

The particular choice of parameters is model-dependent and is made by
trial and error. In the model we consider here rather small value
$T_M=3$ is sufficient to reach linear regime starting from the
vicinity of the sphaleron configuration. Fast linearization is the
advantage of high-dimensional models \cite{Rebbi}
and is due to the volume factor
(note $1/r^2$ in the interaction term in \eq{action2}). This effect is
absent in two-dimensional models.  Moreover, in $-\lambda\phi^4$ model, the
value $T_M=3$ is sufficient for linearization at least up to energies $E \sim
3E_{sph}$. This feature differs the above model from other
four-dimensional models we have tried.

Fast linearization allows to take relatively small space size, $L=3$,
which leads to the sufficiently wide spectrum already at $n_x=40$. The
free spectrum for these values of parameters reproduces the continuum
spectrum with reasonable accuracy up to $\omega\sim 15$.

\section{Results and discussion}

The results presented in this this paper were obtained at $L=3$,
$T_M=3$, $n_x=40$ and $n_t$ varying from $200$ to $300$. For
these values of grid parameters we have found about 1500 solutions to
the boundary value problem with different $E$ and $N$. The region of
$\epsilon$--$\nu$ plane covered by the solutions is shown in
Fig.2. The top left corner of the solution  region (point S in Fig.2) is the
sphaleron. It corresponds to $\epsilon=\nu=1$. At this point the
function $F(\epsilon,\nu)$ vanishes and exponential suppression of
probability disappears.

 Towards small energies the solution region is bound by the line of
periodic instantons (the line SP in Fig.2). In the continuum limit
this line would end up at the point $\epsilon=\nu=0$ which corresponds
to the zero energy instanton. However, the spatial size of periodic
instantons rapidly decreases along this line. At the above values of
grid parameters, the instanton size becomes comparable to the lattice
spacing at $\epsilon\approx 0.4$ (the point P in Fig.2).  At this
point the Newton's algorithm stops to converge. Along the line of
periodic instantons, the function $F(\epsilon,\nu)$ monotonically
decreases and reaches the value $F=-0.6 S_{inst}$ at the point P.

The boundary of convergency continues at approximately constant $\nu$
towards higher energy. It is represented by the line PQ in Fig.2.  The
function $F$ is negative along this line.

The third boundary of the solution region is formed by the line $F=0$
(solid line in Fig.2). It starts at the sphaleron and goes towards
higher energy and smaller number of particles. At energy $E\approx
3E_{sph}$ it comes close to the boundary of convergency, which makes
finite the available part of $\epsilon$--$\nu$ plane.  Our data cover
this available region.

In this paper we concentrate on the dependence of the function $F$ on
energy and number of particles. It is most conveniently represented by
the lines of constant $F$ in the $\epsilon$--$\nu$ plane. These lines
can be obtained by interpolation and are shown in Fig.3.  They start
at the line of periodic instantons with
infinite negative slope. The latter can be seen analytically. Indeed,
from eqs.(\ref{EN eqs}) and (\ref{F=}) one immediately derives that
\begin{equation}
\dpar{N}{E}\Bigm|_F = -{2T\over \theta}.
\label{dNdE}
\end{equation}
The slope $\partial N/\partial E|_F$ becomes infinite at the
periodic instantons where $\theta=0$.

The function $F(\epsilon,\nu)$ determines the maximum probability of induced
false vacuum decay among $N$-particle initial states. The line $F=0$
separates the classically forbidden and classically allowed
regions. We see from Fig.3 that for each energy $E>E_{sph}$ there
exists a minimum number of particles $N_{crit}(E)$ for which the decay
may occur classically without exponential suppression. At
$N>N_{crit}$ one thus may expect the existence of classical
configurations which pass above the barrier.  On the other hand, at
$N<N_{crit}$ all classical solutions bounce off the barrier. By
continuity, when $N$ approaches $N_{crit}$ from above, the classical
solutions which pass above the barrier should spend more and more time
on its top oscillating above the sphaleron, so that in the limit
$N=N_{crit}$ this time goes to infinity.

At $N<N_{crit}$, the decay is a tunneling event which is described by
corresponding solution to our boundary value problem. As discussed
above, the real time part of this tunneling solution represents the
classical evolution of the system after barrier penetration. In our
model this evolution leads to the singularity (point P in Fig.1). In
between two singularities on the real time axis, the tunneling
solution represents the field which comes from infinity, bounces off
the barrier and goes back to infinity. When $N$ approaches $N_{crit}$
from below, the distance between two singularities grows (see Fig.4)
and the solution spends more and more time around the
sphaleron. Although due to instabilities we were not able to trace the
positions of singularities when the distance between them becomes of
order one, we expect that at the line $F=0$ this distance goes to
infinity and the real time part of the tunneling solution reproduces
the infinitely oscillating classical solution described above. Thus,
the configurations which correspond to the line $F=0$ should be
classical solutions infinitely oscillating above the sphaleron.

Due to the existence of minimum number of particles at fixed energy
$E$, one can try to obtain the line $F=0$ by minimizing $N$ over the
set of classical solutions which pass above the barrier.  This
strategy was used in ref.\cite{Rebbi} to set bounds on the forbidden
region. We have tried to use this strategy to reproduce the line $F=0$
found by the tunneling approach. We have obtained bounds which are
noticeably higher than the real position of this line. The reason could
be that the classical solutions which correspond to the line $F=0$ are
infinitely oscillating around the sphaleron and thus are singular
points of the space of solutions which pass above the barrier.

As one can see from Fig.3, the size of allowed region increases
towards higher energy, while the minimum number of particles for which
the decay is not exponentially suppressed, decreases. We have traced
this behaviour up to $E\sim 3E_{sph}$ where the absence of convergency
prevents us from going further with the same grid parameters. Since
the probability $\sigma(E,N)$ is an upper bound for the two-particle
cross section at energy $E$, we conclude that the latter cross section
is exponentially suppressed at $E<3E_{sph}$. Moreover, extrapolating
the behaviour of $N_{crit}(E)$ and assuming that the derivative
$\partial N_{crit}(E)/\partial E$ continues to decrease in absolute
value, one can see that the line $F=0$ does not cross the $N=0$ axis
at least up to $E\sim 10 E_{sph}$. Thus, we conclude that exponential
suppression of the two-particle cross section persists at least up to
$E\sim 10 E_{sph}$.

Careful look at the data shows that the behaviour of the function
$F(\epsilon,\nu)$ may qualitatively change at substantially lower
energy. To see this consider the slope of the lines of constant $F$ as
a function of $E$. The derivative $\partial N/\partial E|_F$ can be
expressed through the known parameters according to \eq{dNdE}. Its
behaviour with energy is shown in Fig.5.  Extrapolating, one can see
that it would become zero at $E_*\approx 3.5 E_{sph}$ and
$N_*\approx 0.35 N_{sph}$. If this indeed happens, then starting at
this energy the line $F=0$ in coordinates $E$ and $N$ would turn up
($d N_{crit}(E)/d E$ would become positive), while the
function $F(E,N_*)$ at $E>E_*$ would decrease with energy.  Clearly,
this behaviour is unphysical.  Indeed, it is always possible to put
the excess energy $E-E_*$ in one quantum particle and thus effectively
decrease the energy involved in the semiclassical tunneling, leaving
the number of particles practically unchanged. The corresponding
initial state would not be semiclassical, and its probability of
tunneling would be $F(E_*,N_*)$, i.e. higher than follows from the
semiclassical picture.  If this situation realizes, it is natural to
expect that starting at $E=E_*$ the correct function $F(E,N_*)$ stays
constant and thus the line $F=0$ never crosses the $N=0$ axis.
The latter would mean that
the two-particle cross section is always exponentially suppressed.

As is clear from Fig.3, in order to go to higher energies and verify
the above conclusions based on extrapolation, one has to extend the
convergency region towards smaller number of particles.  The breakdown
of convergency is most likely related to the lattice spacing in $r$
direction. We have checked that doubling $n_x$ at fixed $L$ extends
the convergency region for periodic instanton solutions.  We expect
that the same effect persists at $\theta\neq 0$. Thus, increasing
$n_x$ could be the way to extend the convergency region and
improve our results. This, however, will require noticeably  more
computer resources.

In conclusion, we would like to stress again the analogy, as long as
instanton-like transitions are concerned, between the $-\lambda\phi^4$
model and the bosonic sector of the Electroweak Theory. On the basis
of this analogy, we expect that the Electroweak Theory is also a suitable
model for studying instanton-like transitions numerically. Since,
because of larger number of fields,
calculations there are at least $4^3$ times slower, they would
require the use of computers much faster than a typical
workstation.

\section*{Acknowledgments}

The authors would like to thank V.A.Rubakov and D.T.Son for numerous
and fruitful discussions at different stages of this work. The work is
supported by ISF grant \#MKT300 and  INTAS grant \#INTAS-94-2352.
The work of P.T. is supported in part by Russian Foundation for
Fundamental Research, grant \#93-02-3812.

\newpage

\section*{Figure captions}

\noindent
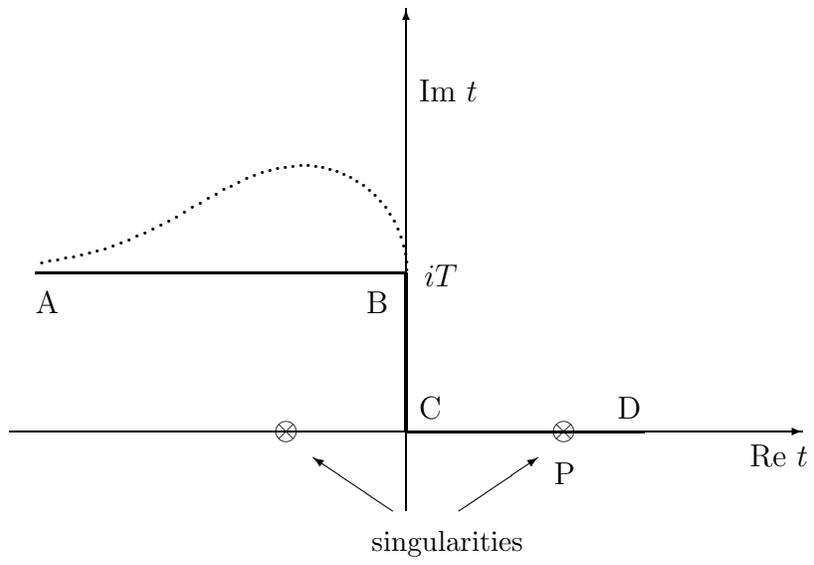
{\bf Fig.1.}~The contour ABCD in the complex time plane where the boundary
value problem is formulated. Crossed circles represent singularities of the
field. Dotted line schematically shows the deformed contour used in
numerical calculations.
\vspace{0.5cm}

\noindent
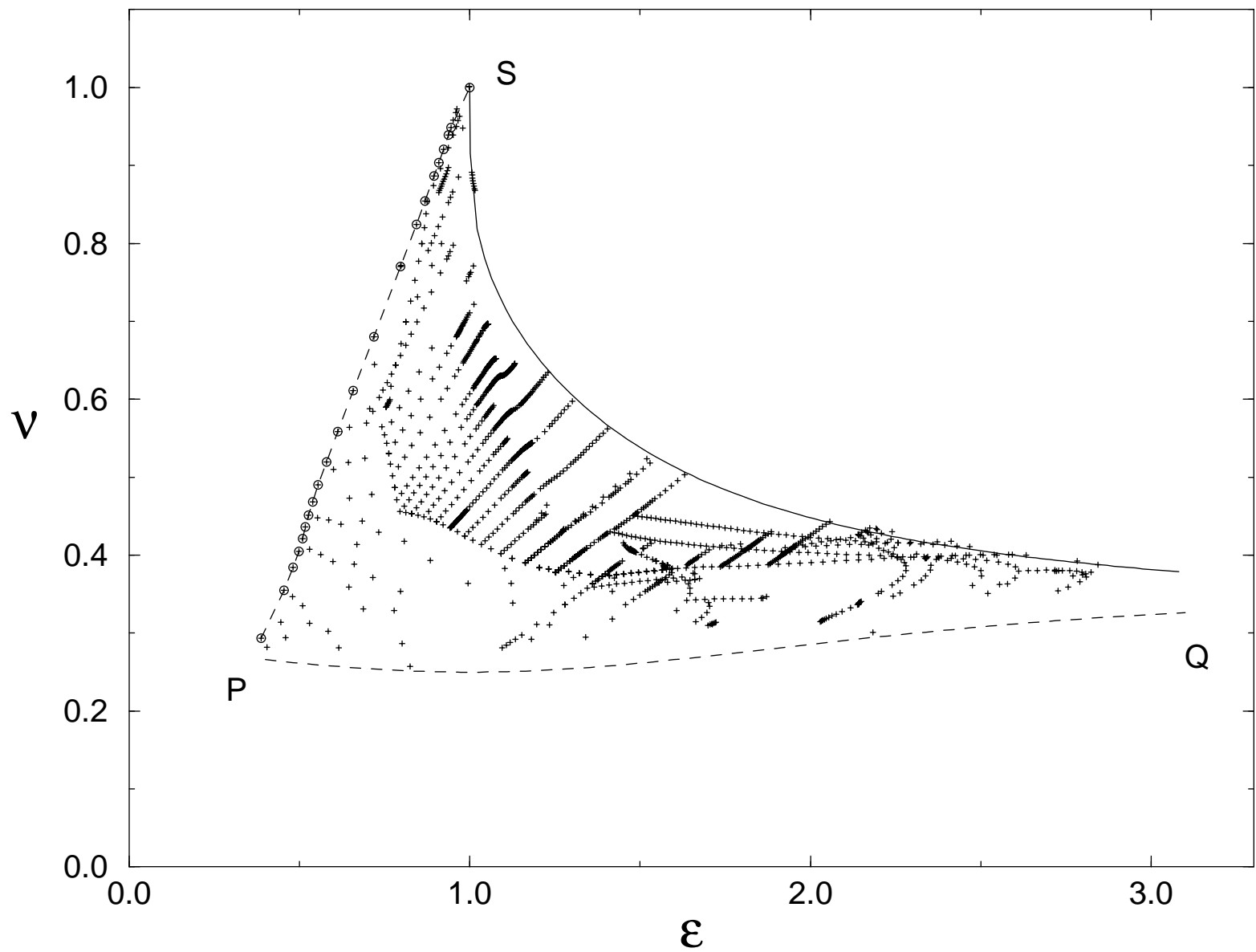
{\bf Fig.2.}~The $\epsilon$--$\nu$ map of the obtained solutions. The point S
corresponds to the sphaleron. Crossed circles lying on the line SP are
periodic instantons. The line PQ represents the boundary of
convergency region, while the solid line corresponds to
\vspace{0.5cm}
$F(\epsilon,\nu)=0$.

\noindent
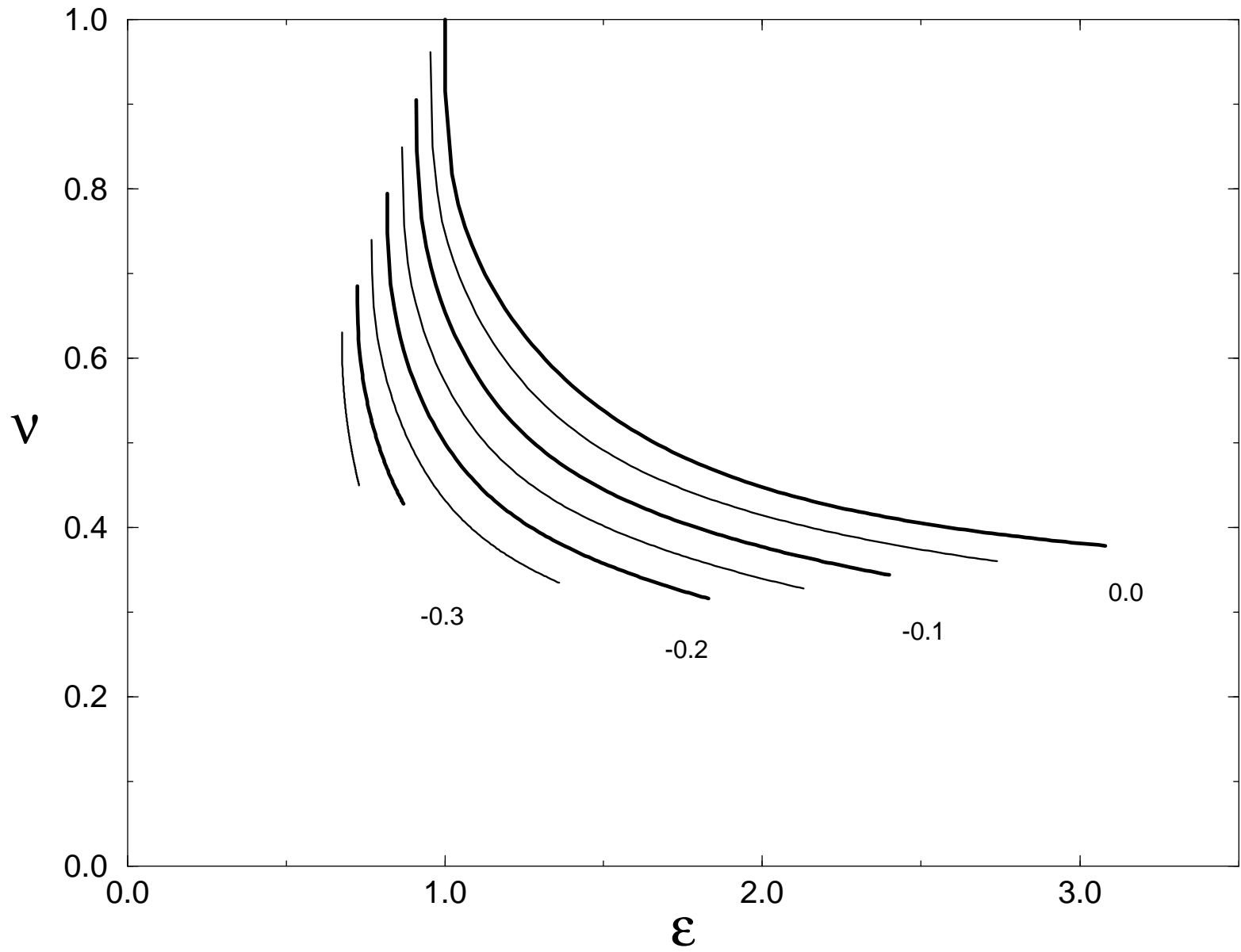
{\bf Fig.3.}~The lines of constant $F(\epsilon,\nu)$ in the
$\epsilon$--$\nu$ plane. Numbers show the value of $F$ in the
units of $S_{inst}$.
\vspace{0.5cm}

\noindent
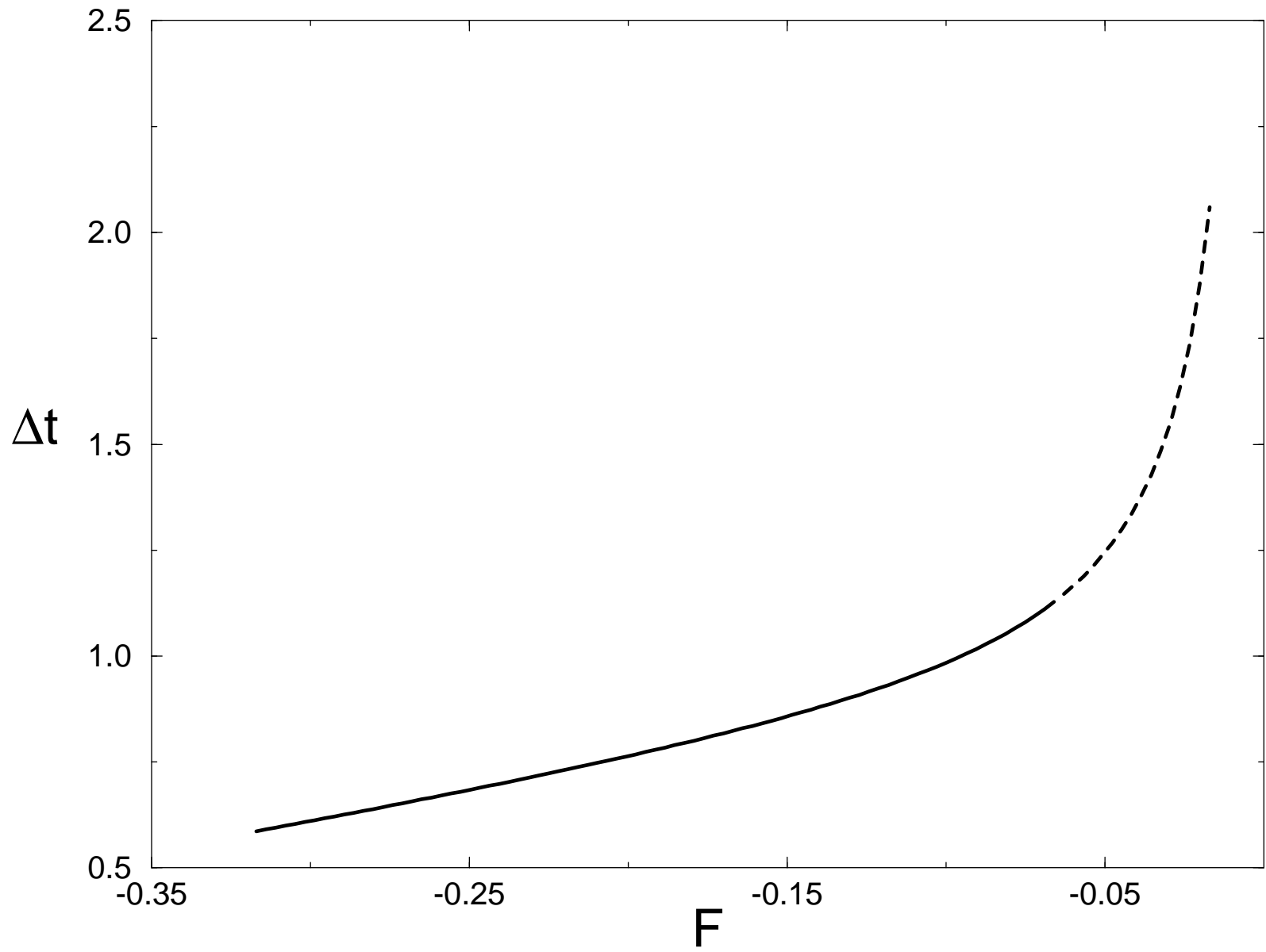
{\bf Fig.4.}~The distance between singularities $\Delta t$ as a
function of $F$ when $F$ approaches zero. Dashed line corresponds to
the region where the procedure of calculating $\Delta t$
is not reliable.
\vspace{0.5cm}

\noindent
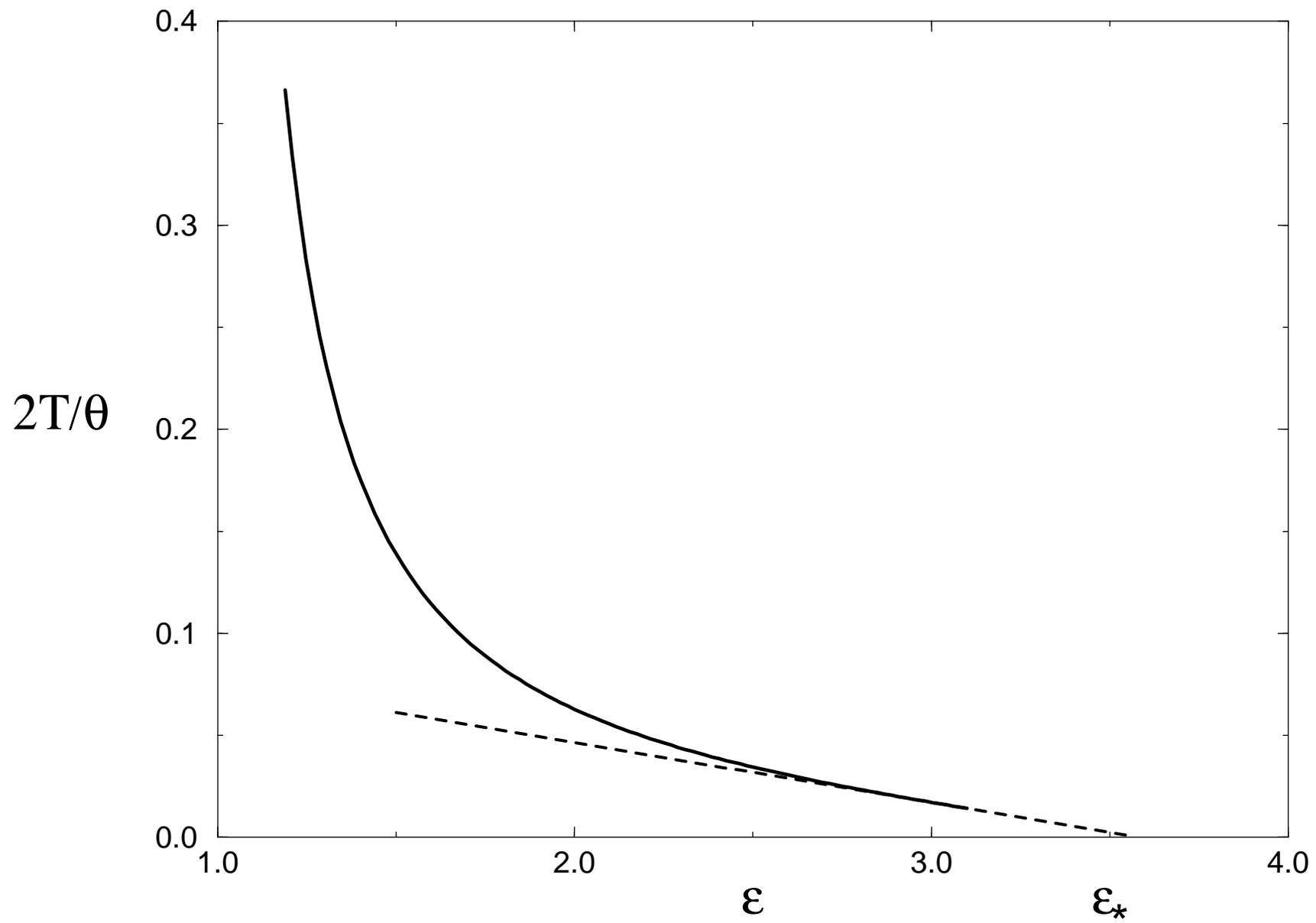
{\bf Fig.5.}~The dependence of $-\partial N/\partial E|_F = 2
T/\theta$ on $\epsilon$. Dashed line represents the extrapolation to
higher energies. $\epsilon_* = E_*/E_{sph}$ corresponds to the point
where the behaviour of $F(\epsilon,\nu)$ may qualitatively change.

\newpage

\iftrue
\begin{picture}(300,250) (-30,0)
\put (0,100){\vector(1,0){300}}
\put (150,70){\vector(0,1){190}}
\thicklines
\put (10,160)  {\line(1,0){140}}
\put (150,160) {\line(0,-1){60}}
\put (150,100) {\line(1,0){90}}
\put  ( 11.99,   163.45)  {\hbox{\hskip -0.11em .}}
\put  ( 14.99,   163.90)  {\hbox{\hskip -0.11em .}}
\put  ( 17.99,   164.39)  {\hbox{\hskip -0.11em .}}
\put  ( 20.99,   164.94)  {\hbox{\hskip -0.11em .}}
\put  ( 23.99,   165.56)  {\hbox{\hskip -0.11em .}}
\put  ( 26.99,   166.24)  {\hbox{\hskip -0.11em .}}
\put  ( 29.99,   166.99)  {\hbox{\hskip -0.11em .}}
\put  ( 32.99,   167.82)  {\hbox{\hskip -0.11em .}}
\put  ( 35.99,   168.74)  {\hbox{\hskip -0.11em .}}
\put  ( 38.99,   169.73)  {\hbox{\hskip -0.11em .}}
\put  ( 41.99,   170.82)  {\hbox{\hskip -0.11em .}}
\put  ( 44.99,   172.00)  {\hbox{\hskip -0.11em .}}
\put  ( 47.99,   173.26)  {\hbox{\hskip -0.11em .}}
\put  ( 50.99,   174.62)  {\hbox{\hskip -0.11em .}}
\put  ( 53.99,   176.05)  {\hbox{\hskip -0.11em .}}
\put  ( 56.98,   177.57)  {\hbox{\hskip -0.11em .}}
\put  ( 59.98,   179.16)  {\hbox{\hskip -0.11em .}}
\put  ( 62.98,   180.80)  {\hbox{\hskip -0.11em .}}
\put  ( 65.97,   182.49)  {\hbox{\hskip -0.11em .}}
\put  ( 68.96,   184.21)  {\hbox{\hskip -0.11em .}}
\put  ( 71.95,   185.94)  {\hbox{\hskip -0.11em .}}
\put  ( 74.94,   187.65)  {\hbox{\hskip -0.11em .}}
\put  ( 77.92,   189.34)  {\hbox{\hskip -0.11em .}}
\put  ( 80.88,   190.95)  {\hbox{\hskip -0.11em .}}
\put  ( 83.61,   192.38)  {\hbox{\hskip -0.11em .}}
\put  ( 86.57,   193.84)  {\hbox{\hskip -0.11em .}}
\put  ( 89.54,   195.18)  {\hbox{\hskip -0.11em .}}
\put  ( 92.49,   196.39)  {\hbox{\hskip -0.11em .}}
\put  ( 95.43,   197.45)  {\hbox{\hskip -0.11em .}}
\put  ( 98.36,   198.34)  {\hbox{\hskip -0.11em .}}
\put  (101.27,   199.05)  {\hbox{\hskip -0.11em .}}
\put  (104.17,   199.58)  {\hbox{\hskip -0.11em .}}
\put  (107.04,   199.92)  {\hbox{\hskip -0.11em .}}
\put  (109.90,   200.07)  {\hbox{\hskip -0.11em .}}
\put  (112.72,   200.02)  {\hbox{\hskip -0.11em .}}
\put  (115.52,   199.77)  {\hbox{\hskip -0.11em .}}
\put  (118.27,   199.32)  {\hbox{\hskip -0.11em .}}
\put  (120.99,   198.67)  {\hbox{\hskip -0.11em .}}
\put  (123.65,   197.83)  {\hbox{\hskip -0.11em .}}
\put  (126.26,   196.80)  {\hbox{\hskip -0.11em .}}
\put  (128.80,   195.57)  {\hbox{\hskip -0.11em .}}
\put  (131.27,   194.15)  {\hbox{\hskip -0.11em .}}
\put  (133.65,   192.53)  {\hbox{\hskip -0.11em .}}
\put  (135.93,   190.73)  {\hbox{\hskip -0.11em .}}
\put  (138.10,   188.74)  {\hbox{\hskip -0.11em .}}
\put  (140.13,   186.57)  {\hbox{\hskip -0.11em .}}
\put  (142.03,   184.22)  {\hbox{\hskip -0.11em .}}
\put  (143.77,   181.69)  {\hbox{\hskip -0.11em .}}
\put  (145.33,   178.98)  {\hbox{\hskip -0.11em .}}
\put  (146.70,   176.10)  {\hbox{\hskip -0.11em .}}
\put  (147.87,   173.04)  {\hbox{\hskip -0.11em .}}
\put  (148.82,   169.8)   {\hbox{\hskip -0.11em .}}
\put  (149.63,   166.8)   {\hbox{\hskip -0.11em .}}
\put  (149.98,   163.8)   {\hbox{\hskip -0.11em .}}
\put  (150.00,   160.8)   {\hbox{\hskip -0.11em .}}
\put  (100,97) {$\otimes$}
\put  (205,97) {$\otimes$}
\thinlines
\put  (137,55) {\small singularities}
\put  (145,70) {\vector(-3,2){30}}
\put  (170,70) {\vector(3,2){30}}
\put  (10,145)  {A}
\put (135,145) {B}
\put (155,105) {C}
\put (230,105) {D}
\put (206,80) {P}
\put (157,155) {$iT$}
\put (155,225) {Im~$t$}
\put (280,87)  {Re~$t$}
\put (140,10)   {Fig.1}
\end{picture}
\fi

\end{document}